\documentclass[prl,twocolumn,showpacs,amsmath,amssymb]{revtex4}

\usepackage{graphicx}%Include figure files
\usepackage{dcolumn}%Align table columns on decimal point
\usepackage{bm}% bold math
\usepackage{epsfig}

\begin{document}

\title{Entanglement of Electron Spin and Orbital States in Spintronic Quantum Transport}

\author{Branislav K. Nikoli\' c}
\affiliation{Department of Physics and Astronomy, University
of Delaware, Newark, DE 19716-2570}

\begin{abstract}
An electron within a mesoscopic (quantum-coherent) spintronic structure is
described by a single wave function which, in the presence of both
charge scattering and spin-orbit coupling, encodes an information
about {\em entanglement} of its spin and orbital degrees of
freedom. The quantum state---an {\em improper} mixture---of experimentally 
detectable spin subsystem is elucidated by evaluating quantum information theory 
measures of entanglement  in the scattering states which determine {\em quantum transport} 
properties  of spin-polarized electrons injected  into a two-dimensional disordered Rashba 
spin-split conductor that is attached to the ferromagnetic source and drain electrodes. 
Thus, the Landauer transmission matrix, traditionally evaluated to obtain the spin-resolved 
conductances, also yields the reduced spin  density operator allowing us to extract 
quantum-mechanical measures of the detected electron spin-polarization and spin-coherence, thereby 
pointing out how to avoid detrimental {\em decoherence} effects on spin-encoded 
information transport through semiconductor spintronic devices. 
\end{abstract}


\maketitle

Recently, a  considerable effort has been devoted to gain control
over a  ``neglected'' property of an electron---its spin---and
demonstrate how it can be exploited  as a carrier of
information~\cite{spintronics}. This field of research, dubbed
{\em spintronics}, is envisioned to bring new paradigms for both
classical and quantum information processing. Manipulation of
spins as possible carriers of classical information, when combined
with traditional electronics manipulating the charge, offers
exciting possibilities for novel devices that would process,
store, and communicate information on the same
chip~\cite{datta90}. Furthermore, phase-coherent dynamics of
spin-$\frac{1}{2}$ as a generic two-level system has been one of
the most natural candidates for a qubit in quantum computers~\cite{qc}. The 
attempts to utilize long-lived (because of the weak coupling to environmental
forces) spin quantum states in semiconductors, either as
spin-polarized currents or solid state qubits, is currently at the 
basic research phase where various conceptual problems are to be surmounted before functional
devices can be transferred to an engineering phase. Besides
pursuits of efficient room temperature injection~\cite{injection}
from a ferromagnetic source (metallic or semiconducting) into a
non-magnetic semiconductor, or tunability of spin-dependent
interactions~\cite{nitta}, some well-known concepts of the spin
physics are to be reexamined in this new guise. For instance, one
would like to know the fate of the spin-polarization of injected
electrons (which can change its properties or diminish altogether
due to exposure to various controlled~\cite{nitta,datta90} or
uncontrolled~\cite{jaro,qc} spin-dependent interactions,
respectively) in the course of transport through complicated
solid-state environment, thereby building a firm
quantum-mechanical ground for the understanding of what is
actually measured in the final stage of experiments in 
spintronics~\cite{hammar}.
\begin{figure}
\centerline{\psfig{file=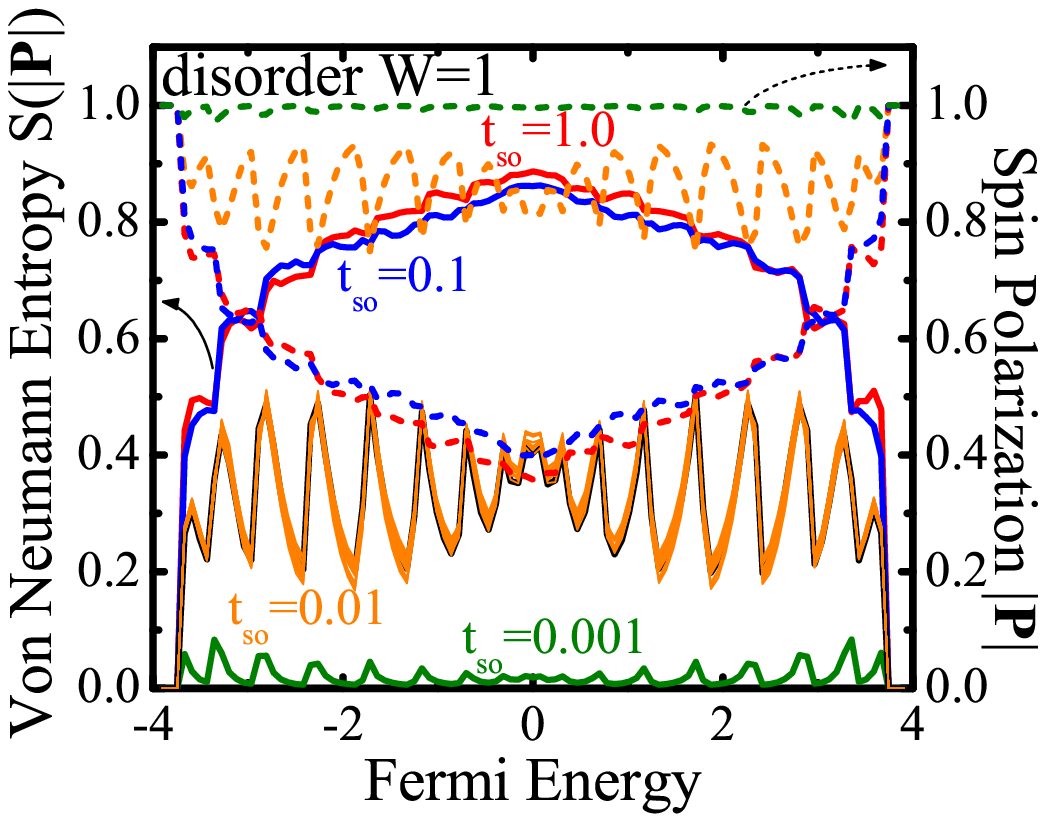,height=2.5in,angle=0} }
\vspace{-0.15in}
\centerline{\psfig{file=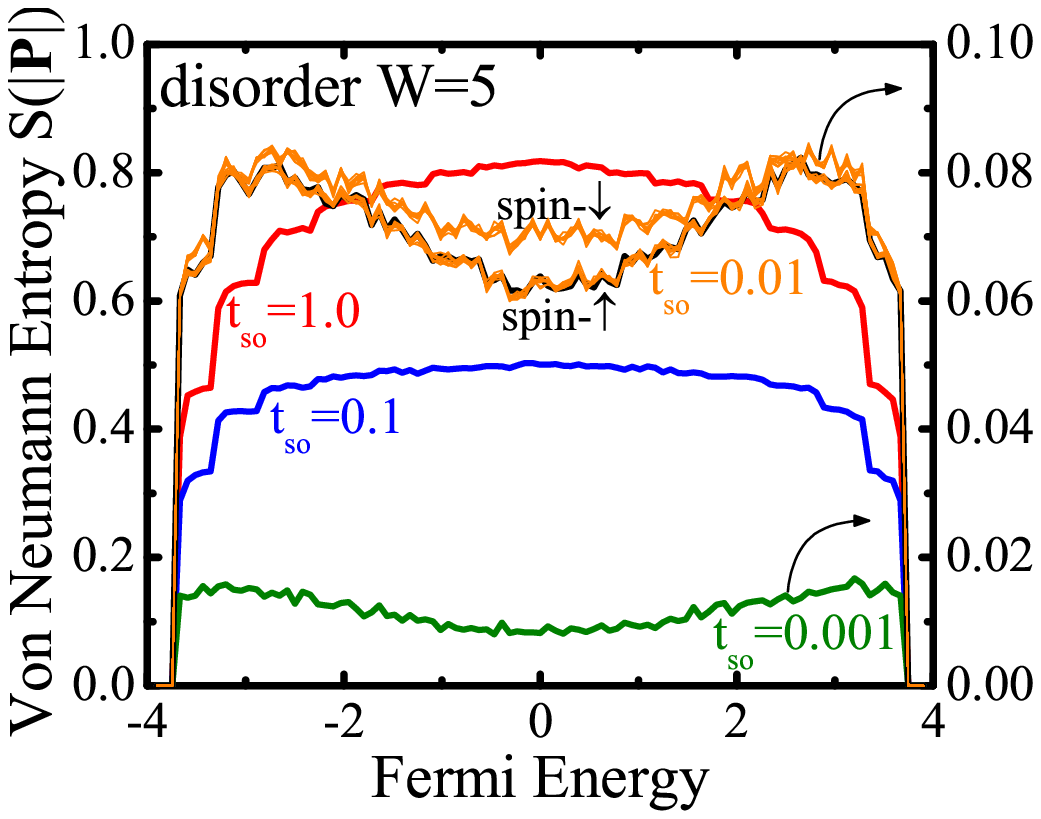,height=2.5in,angle=0} }
\caption{The von Neumann entropy ({\em solid}) and polarization ({\em dashed}) 
of the spin quantum state of a detected electron in  the right lead 
of a Ferromagnet-Semiconductor-Ferromagnet structure where
spin-$\uparrow$ electron is injected from the left lead into the
central region (modeled on a tight-binding lattice $100 \times
10$) with Rashba SO interaction and disorder (disorder-averaging is 
performed over 10000 samples) characterized by the
strengths $t_{\rm so}=\alpha/2a$ and $W$, respectively. The
orbital state of the injected electron is chosen to correspond to
the first (the lowest subband) conducting channel in the Landauer scattering picture of transport, 
except for $t_{\rm so}=0.01$ entropy curves that depict all of the twenty 
spin-polarized channels, exhibiting similar features for each ($\uparrow$ or 
$\downarrow$) spin species.} \label{fig:entropy}
\end{figure}

The essential features of  spin dynamics are captured by two
key time scales: spin relaxation $T_1$ and spin decoherence $T_2$
time~\cite{kikkawa}. The time $T_1$ is classical in nature (i.e.,
it does not involve any superpositions of quantum states) since it
determines lifetime of an excited spin state (aligned along the
external magnetic field)~\cite{jaro}---as studied since 1950s by,
e.g., exciting a nonequilibrium population of spin-polarized
electrons in the skin depth of a metal, where microwave radiation
is absorbed in an electron spin-resonance experiment and the
diffusion of spins into the bulk is then traced~\cite{early_spin}. On 
the other hand, the spin decoherence time $T_2$ has received increased 
attention in quantum engineering of single spins where during $T_2$ relative phase in
the  superpositions of $| \!\! \uparrow\rangle$ and $|\! \!
\downarrow \rangle$ spin quantum states is well-defined~\cite{kikkawa}. Thus,
long $T_2$ ensures enough time for quantum coherence in 
quantum computing with spintronic qubits~\cite{qc} or in exploiting 
spin interference effects in the envisaged semiconductor spintronic devices ($T_2$
can reach $100$ ns thereby allowing for the transport of coherent
spin states over length scales $L_2 < 100$ $\mu$m~\cite{kikkawa}).  The principal modes of 
decoherence and relaxation are exchange coupling  with nuclear or other electronic spins, as well 
as the spin-orbit (SO) coupling to impurity atoms and defects~\cite{jaro,kikkawa}.

While paramount problems in spin injection~\cite{injection} are currently
under scrutiny, efficient detection of spin in solid state systems, whose
weak coupling to the environment serves as an impediment here, remains an equal
challenge~\cite{hammar}. Surprisingly enough, there are many different notions 
of spin-polarization of electrons detected via transport measurements in 
metallic~\cite{jedema}  or semiconductor~\cite{parek} spintronic structures. 
On the other hand, strict quantum-mechanical description of both spin-polarization 
and spin-coherence is unique: the quantum state  of a spin subsystem has to be described by a 
{\em spin density operator}. The spin density operator  $\hat{\rho}_s$ makes it possible to  
predict  the result of a measurement of any quantum-mechanical observable related to spin 
only, thereby accounting for both transport~\cite{hammar} and optical~\cite{spintronics,kikkawa,pure_spin} 
(which are particularly relevant for detection of pure spin currents that are not accompanied by a net charge 
current~\cite{pure_spin}) detection schemes.  Here we demonstrate that spin-dependent Landauer transmission 
matrix~\cite{nikolic}, as one of the  basic tools of quantum transport theory~\cite{carlo_rmt}, also 
contains the information about $\hat{\rho}_s$   of detected electrons. Furthermore, 
extracting $\hat{\rho}_s$ follows the same route pursued in quantum information science when 
studying  entanglement in bipartite quantum systems~\cite{galindo}. As an example of the quantum 
information like analysis of a spin state of electrons comprising a current in spintronic nanostructures, 
Fig.~\ref{fig:entropy} plots the von Neumann entropy $S(\hat{\rho}_s)$ and corresponding spin-polarization 
$|{\bf P}|$ determined by $\hat{\rho}_s$ of electrons that are collected in the left ferromagnetic 
lead after spin-polarized electrons have been injected from the right lead 
and {\em transported} phase-coherently through a two-dimensional disordered 
Rashba-split electron gas. The spin-splitting of the conduction band occurs 
due to the Rashba~\cite{rashba} type of a SO interaction (which always
act like an effective momentum dependent magnetic field in the rest frame of an 
electron). This framework provides direct insight into possibility to transport 
quantum or classical information which would be encoded in coherent superpositions of spin state, as 
envisaged in spin qubits or Datta-Das proposal for spin-field effect transistor (spin-FET)~\cite{datta90}, 
respectively: $S > 0$ (or, equivalently, $|{\bf P}|<1$) signals that purity of spin quantum states is lost 
and coherent quantum-mechanical superposition of states is impeded. To elucidate the meaning of these findings, the rest of the paper  
introduces a technique to quantify  {\em ``nonequilibrium'' entanglement} of spin and orbital degrees of freedom 
and spin-polarization  of the scattering quantum states, which are traditionally studied in  mesoscopic physics 
to extract transport quantities (such as conductance, shot noise, \ldots) through the Landauer-B\" utiker formalism~\cite{carlo_rmt}.

The electron spin as physical observable is described by an
operator $\hat{\bm S} = \hbar \hat{\bm{\sigma}}/2$
[$\hat{\bm{\sigma}}=(\hat{\sigma}_x,\hat{\sigma}_y,
\hat{\sigma}_z)$ is a vector of the Pauli spin matrices] acting in
a two-dimensional vector space ${\mathcal H}_{s}$. If spin state
is accounted for by a vector in this two-dimensional space $|\sigma
\rangle \in {\mathcal H}_{s}$, then it is by definition fully
polarized along, e.g., the $z$-axis $\hat{\sigma}_z |\sigma\rangle
=\pm |\sigma\rangle$ ($\sigma = \uparrow, \downarrow$ is a quantum
number labeling eigenvectors with +1 or -1 eigenvalue,
respectively). Another possibility exists in quantum mechanics:
the spin state is not as pure as $|\sigma \rangle$, but  rather it
is described by a density operator, e.g., $\hat{\rho}_s  =
\rho_{\uparrow \uparrow}|\!\! \uparrow\rangle \langle \uparrow \!\!| +
\rho_{\downarrow \downarrow}|\!\!\downarrow \rangle \langle \downarrow \!\!
|$. The most general description of a two-level system, which is in
the statistical superposition of two state vectors, is
accomplished by a density operator~\cite{ballentine}
\begin{equation}
\label{eq:rhos_general} \hat{\rho}_s  =
\frac{1+{\bf P} \cdot \hat{\bm{\sigma}}}{2}.
\end{equation}
The vector ${\bf P}$ quantifying the ``purity'' of the quantum state of also 
provides a strict quantum-mechanical notion of spin-polarization $|{\bf P}|$ 
for spintronic experiments. For example: (i) $|{\bf P}| = 1$ stands for a {\em pure} quantum state; (ii) $|{\bf P}| = 0$ stands for a 
non-pure state ({\em mixture} or statistical superposition) that 
is completely unpolarized; (iii) for intermediate values $0 < |{\bf P}| < 1$,
spin-$\frac{1}{2}$ particle is partially spin-polarized mixture. 
Any pure quantum state can be viewed as a  coherent superposition
$|\Sigma \rangle = \cos (\theta/2) |\!\!\uparrow \rangle +
e^{i\phi} \sin (\theta/2) |\!\!\downarrow\rangle$,  which, being
an eigenvector of $\hat{\bm{\sigma}} \cdot {\bf P}$, is fully
polarized in the direction ${\bf P}=(\theta,\phi)$ specified in
terms  of the angles $\theta$ and $\phi$ on a Bloch
sphere~\cite{galindo}. In this formal language, the dynamics of
spin decoherence and spin relaxation follows a simple `chain of
events': a pure state $|\Sigma \rangle \langle \Sigma|= \left(
\begin{array}{cc} \cos^2 \frac{\theta}{2} &
\frac{e^{-i\phi}\sin \theta   }{2}   \\
\frac{e^{i\phi}\sin \theta  }{2} & \sin^2 \frac{\theta}{2}
         \end{array} \right)$ decoheres into $\left( \begin{array}{cc}
        \cos^2 \frac{\theta}{2}  & 0   \\
       0 & \sin^2 \frac{\theta}{2} \\
         \end{array} \right)$, on the time scale $T_2$ due to
         {\em entanglement} to the environment (such vanishing of
the diagonal elements of a density operator is the general
phenomenology of any decoherence process~\cite{zeh});  the
remaining diagonal elements equilibrate on the scale $T_1$
yielding finally  $\left( \begin{array}{cc}
       \rho^{\rm eq}_\uparrow & 0  \\
        0 & \rho^{\rm eq}_\downarrow \\
         \end{array} \right)$ (typically $T_2 < T_1$)~\cite{galindo}.

The spin-polarized electron injected into quantum-coherent spintronic structure 
remains in a pure state $|\Psi\rangle$, which is a principal feature of mesoscopic 
system studied over the past two decades by fabricating nanoscale samples and 
measuring transport at low enough temperatures ensuring that the sample
size $L < L_\phi$  is smaller than the dephasing length $L_\phi$ (determined by 
inelastic processes)~\cite{carlo_rmt}. However, traditional mesoscopic experiments 
probe the quantum coherence of orbital wave functions, where on the scale $L_\phi$ 
one observes quantum interference effects due to the preservation of relative phase in the
superpositions of spatial states of an electron. On the other hand, vector $|\Psi\rangle$ 
is a pure state in the tensor product of the orbital ${\mathcal H}_o$  and the spin Hilbert 
space, $|\Psi\rangle \in {\mathcal H} = {\mathcal H}_o \otimes {\mathcal
H}_s$. In general, this state does not have to be separable,
meaning that it cannot be decomposed into a tensor product of two
vectors  $|\Phi\rangle \otimes |\sigma \rangle$ (unless the orbital
and spin coherence, i.e., $L_\phi$ and $L_2$, are completely
independent~\cite{qc}). Instead, $|\Psi\rangle$ is a superposition of
uncorrelated states $|\phi_\alpha\rangle \otimes |\sigma\rangle$
comprising a basis in ${\mathcal H}$: $|\Psi\rangle =
\sum_{\alpha,\sigma} c_{\alpha,\sigma} |\phi_\alpha\rangle \otimes
|\sigma\rangle$, where $|\phi_\alpha \rangle \in {\mathcal H}_o$
is a  basis in the orbital factor space ${\mathcal H}_o$. Thus,
when the interplay of SO interaction and scattering at
impurities, boundaries or interfaces takes place, orbital and spin
quantum subsystems become {\em entangled}. Since their individuality
is lost, they cannot be assigned a pure state vector any more.
Nevertheless, the result of a measurement of any spin observable,
described by an operator $\hat{O}_s$ acting  in  ${\mathcal H}_s$,
is accounted by $\langle O_s \rangle = {\rm Tr}_s \, [\hat{\rho}_s
\hat{O}_s$]. Here the reduced density operator $\hat{\rho}_s = {\rm
Tr}_o \hat{\rho}=\sum_\alpha \langle \phi_\alpha| \hat{\rho}|
\phi_\alpha \rangle$ is obtained by partial tracing, over the
orbital degrees of freedom, of the pure state density operator
$\hat{\rho}=|\Psi\rangle \langle \Psi|$. Equivalently, $\hat{\rho}_o={\rm Tr}_s \, \hat{\rho}$ plays
the same role in the measurement of spatial observables. However, the
knowledge of $\hat{\rho}_s$ and $\hat{\rho}_o$ does not exhaust all
non-classical correlation between electron orbital and spin
degrees of freedom embodied in $|\Psi \rangle$, i.e., 
$\hat{\rho} \neq \hat{\rho}_o \otimes \hat{\rho}_s$  except when the full state is 
separable $|\Psi \rangle =|\Phi\rangle \otimes |\sigma\rangle$.

These concepts are usually  discussed in the context
of composite systems of two particles (like the notorious 
Einstein-Podolsky-Rosen states violating the Bell 
inequalities~\cite{ballentine}). In fact, recent diffusion of the
ideas of quantum information theory has led to reexamination~\cite{nielsen} 
of many-body correlations in quantum systems by quantifying the 
entanglement in their wave functions (which are available  in condensed matter 
physics only in a limited number of exactly or variationally solvable models). 
Nonetheless,  the formalism is the same for any composite quantum system, 
regardless of whether the  subsystems are particles or additional {\em internal} 
degrees of freedom---the joint Hilbert space ${\mathcal H}$ of a
multipartite system is always constructed as a tensor product of the
Hilbert spaces of quantum subsystems. The entanglement between
subsystems is manifestation of linear superpositions
of states in ${\mathcal H}$. It can be quantified using the von
Neumann entropy of a density operator, e.g., in the case of 
bipartite systems $S(\hat{\rho}_s)=-{\rm Tr}\, [\hat{\rho}_s \log_2
\hat{\rho}_s]=S(\hat{\rho}_o)$. Essentially, any measure of the
entanglement in the pure bipartite state is just a function of the
eigenvalues of $\hat{\rho}_s$ or $\hat{\rho}_o$ (quantifying
entanglement of pure-multipartite or mixed-bipartite states is
far more intricate problem~\cite{galindo}). Evidently, the entropy 
of a pure state is zero [e.g., for fully-polarized spin $S(|\Sigma\rangle 
\langle \Sigma|)=0$], and becomes positive for mixtures.

It is insightful to recall a fundamental conceptual difference  between the 
mixed states of  entangled subsystems, the so-called {\em improper mixtures}~\cite{landau}, 
and  proper mixtures stemming from subjective lack of knowledge when  
insufficient filtering of an ensemble takes place during a quantum state 
preparation~\cite{ballentine}. Textbook examples of {\em proper mixtures} 
are: conventional unpolarized currents, where $\rho_{\uparrow \uparrow}=\rho_{\downarrow \downarrow}=1/2$ 
gives rise to $\hat{\rho}_s=\hat{I}_s/2$; or thermal equilibrium states in 
macroscopic solids~\cite{galindo}. Although one uses the same mathematical 
tool $\hat{\rho}_s$ in Eq.~(\ref{eq:rhos_general}) for both proper and 
improper mixtures, no real statistical ensemble of different spin states 
exists within the conductor that would correspond to improper mixtures 
(the full state is still pure). Injection of partially polarized current, 
i.e., proper mixture, can lead to an entangled mixed bipartite state in 
the opposite lead. 

These formal tools offer a strict quantum-mechanical route toward understanding 
the spin-polarization of electrons in a paradigmatic two-terminal
spintronic device~\cite{datta90}, where  two-dimensional electron
gas (2DEG) in a semiconducting  heterostructure is attached to two 
ferromagnetic leads. The non-magnetic part
of the device can be modeled by a single-particle Hamiltonian~\cite{nikolic}
\begin{eqnarray}\label{eq:hamilton}
\hat{H} & = & \left( \sum_{\bf m} \varepsilon_{\bf
m}|{\bf m} \rangle \langle {\bf m}|  +  t \sum_{\langle {\bf
m},{\bf n} \rangle} |{\bf m} \rangle \langle {\bf n}| \right)
  \otimes \hat{I}_{\rm s} \nonumber \\ && + \frac{ \alpha \hbar}{2a^2 t}
(\hat{v}_x \otimes \hat{\sigma}_y - \hat{v}_y \otimes
\hat{\sigma}_x).
\end{eqnarray}
Here the orbital part ($\hat{I}_{\rm s}$ is a unit operator in
${\mathcal H}_s$) is a standard tight-binding Hamiltonian defined
on a square lattice $N_x \times N_y$ where electron can hop, with
hopping integral $t$ (a unit of energy), between nearest-neighbor
$s$-orbitals $\langle {\bf r}|{\bf m}\rangle =  \psi({\bf r}-{\bf
m})$ located on the lattice sites  ${\bf m}=(m_x,m_y)$. In clean
samples $\varepsilon_{\bf m}$  is constant, while in disordered
ones random potential is simulated by taking $\varepsilon_{\bf m}$
to be uniformly distributed over the interval $[-W/2,W/2]$. The SO
term [$\hat{\bm v}=(\hat{v}_x,\hat{v}_y,\hat{v_z})$ is the
velocity operator], acting effectively as an ``entangler'', is the Rashba~\cite{rashba} 
interaction which arises from the asymmetry along the $z$-axis of the confining
quantum well electric potential that creates a two-dimensional (2D)
electron gas ($xy$-plane) on a narrow-gap semiconductor (such as InAs)
surface. This type of SO coupling, which is different from the more
familiar impurity induced and position dependent ones in metals, is
particularly important for spintronics since its strength $\alpha$ can
be tuned in principle by an external gate electrode~\cite{nitta}. It is, therefore,
envisaged as a tool to control the precession (i.e., unitary
quantum evolution of coherent superpositions of spin states)
of injected spin-polarized electrons  during their ballistic
flight, thereby modulating the current in the opposite lead of the spin-FET~\cite{datta90}.

The spin-polarized electrons are injected from the left
ferromagnetic lead into 2DEG, and then either $\uparrow$- or
$\downarrow$-electrons (or both) are collected in the right lead
(which are semi-infinite in theory). To study entanglement in such
{\em open} mesoscopic system, where electrons can escape from the
sample through the leads, we use the fact that pure states of
injected and collected electrons are linear superpositions of
asymptotic scattering  ``channels'' in the Landauer picture of
quantum transport~\cite{nikolic,carlo_rmt}. However, in
spintronics each channel $|n \sigma\rangle$ (at the Fermi energy
$E_F$ in the leads) is fully spin-polarized~\cite{nikolic} and denotes a
separable state vector $\langle {\bf r} |n \sigma \rangle^\pm =
\Phi_n(y) \otimes \exp(\pm i k_n x) \otimes | \sigma \rangle$, which
is characterized by a real wave number $k_n > 0$, transverse wave
function $\Phi_n(y)$ defined by  quantization of transverse
momentum in the leads of a finite cross section (with hard wall
boundary conditions implemented here), and a spin  factor state
$|\sigma \rangle$. When injected electron is prepared in the
channel $|{\rm in}\rangle \equiv | n,\sigma \rangle$, a pure state
will emerge in the other lead $|{\rm out} \rangle =
\sum_{n^\prime \sigma^\prime} {\bf t}_{n^\prime n,\sigma^\prime
\sigma} |n^\prime \rangle \otimes |\sigma^\prime \rangle$. This
coherent superposition introduces a transmission matrix ${\bf t}$,
which is defined in the basis of incoming spin-polarized channels
of the injecting lead and outgoing channels in the detecting
lead~\cite{carlo_rmt,nikolic}. It encodes a unitary transformation
that the central region performs on the incoming electronic wave
function from the left lead. To compute the ${\bf t}$-matrix
efficiently, we switch from wave functions to a single-particle
real$\otimes$spin space Green function (which is an inverse of the
non-Hermitian operator $\hat{H} + \hat{\Sigma}$, with self-energy
$\hat{\Sigma}$ modeling the open quantum system attached to 
the external leads), as elaborated in more detail in Ref.~\cite{nikolic}.

To each of the outgoing pure states we assign the density
operator as the projector onto that state 
\begin{equation}\label{eq:rho_full}
\hat{\rho}^{n \sigma \rightarrow {\rm out}}=\sum_{n^\prime
n^{\prime\prime} \sigma^\prime \sigma^{\prime\prime}} {\bf
t}_{n^\prime n,\sigma^\prime \sigma} {\bf t}_{n^{\prime
\prime} n,\sigma^{\prime\prime} \sigma}^* |n^\prime \rangle \langle
n^{\prime\prime}| \otimes |\sigma^\prime \rangle \langle
\sigma^{\prime\prime} |,
\end{equation} 
where all matrices here depend on $E_F$. Taking partial trace,
which is technically just the sum of all $2 \times 2$ block
matrices along the diagonal of $\hat{\rho}^{n \sigma \rightarrow {\rm
out}}$, we obtain the {\em exact} reduced density 
operator for the spin subsystem 
\begin{equation}\label{eq:rho_partial}
\hat{\rho}^{n \sigma \rightarrow {\rm out}}_s=\sum_
{n^\prime \sigma^\prime\sigma^{\prime\prime}} {\bf
t}_{n^\prime n,\sigma^\prime \sigma} {\bf
t}_{n^\prime n, \sigma^{\prime\prime} \sigma}^*  |\sigma^\prime \rangle
\langle \sigma^{\prime\prime} |.
\end{equation}
From $\hat{\rho}^{n \sigma \rightarrow {\rm out}}_s$ we can extract: the components of the
polarization vector ${\bf P}=(p_x,p_y,p_z)$ defined by
Eq.~(\ref{eq:rhos_general}), ${\bf P}={\rm Tr}\, [\hat{\rho}_s
\hat{\bm \sigma}]$; and the von Neumann entropy of improper
mixture generated by spin-orbit entanglement, $S(\hat{\rho}_s)=-\frac{1}{2}(1+|{\bf P}|) \log_2
[\frac{1}{2}(1+|{\bf P}|)] - \frac{1}{2}(1-|{\bf P}|) \log_2
[\frac{1}{2}(1-|{\bf P}|)]$. Since the von Neumann entropy (a {\em standard} measure of the 
degree of  entanglement in bipartite pure quantum states~\cite{galindo}) is a function 
of the eigenvalues of a hermitian operator, it is essentially a measurable quantity. Moreover, 
the case of  spin-$\frac{1}{2}$ is rather unique~\cite{ballentine}: the spin density operator 
$\hat{\rho}_s$, the spin polarization $|{\bf P}|$, and the degree of entanglement $S(\hat{\rho}_s) \equiv S(|{\bf P}|)$ are 
all obtained  once the average components of the spin operator $\langle \hbar \hat{\sigma}_x/2 
\rangle$, $\langle \hbar \hat{\sigma}_y/2 \rangle$, $\langle \hbar \hat{\sigma}_z/2 \rangle$ have 
been measured.

We are now fully armed to revisit the results of Fig.~\ref{fig:entropy}
for injection of electrons that are spin-polarized along the direction
of transport (as in the spin-FET proposal~\cite{datta90}), denoted as the $x$-axis here, 
through a mesoscopic conductor modeled on a $100 \times 10$
lattice. Within the conductor electrons experience different strengths of
the disorder $W$ and the Rashba SO coupling $t_{\rm so}=\alpha/2a$ ($a$ is the
lattice spacing). We rely here on the quantum intuition, rather than the ``unreliable'' 
classical one that invokes picture of spin precession whose axis changes
every time the electron momentum is knocked out due to the scattering off
impurities---the so-called Dyakonov-Perel (DP) mechanism~\cite{jaro}.
Such process diminishes electron spin-polarization $|{\bf P}| <1$ by entanglement, 
as signaled by $S>0$ in Fig.~\ref{fig:entropy}, which is sensitive even to small disorder $W=1$  
and negligible $t_{\rm so}=10^{-3}$ [around the band center a
mean free path is roughly $\ell \simeq 30  (t/W)^2$]. 
Moreover, our quantum framework goes beyond classical insight, which predicts that 
spin-diffusion length is independent of the mean free path (the so-called motional narrowing 
effect): further increasing of the disorder impedes the DP decoherence mechanism~\cite{jaro,parek} 
thereby restoring the partial quantum coherence of the mixed state of the spin subsystem. This is 
demonstrated by the decrease of entropy $S_{\rm W=5} < S_{\rm W=1}$ when rather strong 
disorder $W=5$ is introduced into the conductor  with fixed and uniform $t_{\rm so}$. However, 
we see that partial spin-polarization for a very wide range of parameters exhibits an 
oscillatory structure (that is not seen in naive measures of spin-polarization~\cite{parek}) 
as a function of $E_F$ at which the zero-temperature quantum transport occurs. In fact, the 
most interesting case occurs for $t_{\rm so}=0.01$, corresponding to currently achievable strengths of the 
Rashba SO coupling~\cite{nitta}, where polarization reaches $|{\bf P}| \approx
0.95$ at discrete values of $E_F$. To evade spin-charge entanglement,  the spin-FET~\cite{datta90} 
would have to operate in a strictly ballistic transport regime. Nonetheless, Fig.~\ref{fig:entropy} 
suggest that careful crafting of the device parameters makes possible to ``purify'' almost completely 
the quantum state of the  spin subsystems, therefore restoring spin coherence, even when charge  
scattering takes place in the presence of the engineered SO interaction.

In summary, it is shown that quantum information style analysis of spin-orbit 
entanglement in the scattering states of quantum transport theory provides a thorough 
understanding of the state of transported spins through spintronic devices, ultimately 
accounting for any spin detection experimental scheme. Although the amplitudes of the Landauer 
transmission matrix elements directly yield conductance, we demonstrate that analysis of  both 
their phases and amplitudes reveals ``nonequilibrium'' spin-orbit entanglement. The spin   polarization, 
spin coherence, and the degree of entanglement are all  extracted from the  reduced spin density 
operator,  whose dynamics is formulated for nonequilibrium steady transport state without invoking any 
master equations of `open quantum system' approaches~\cite{zeh} (note that the reduced density operator 
no longer obeys the von Neumann equation, whereas the total wave function evolves according to the 
Schr\" odinger equation). The principal insight gained for the future of spin-FET devices is that, due to any scattering mechanism 
(even without disorder, scattering can occur at the interfaces and boundaries, particularly at 
the lead-semiconductor interface due to difference in electronic states with and without the Rasha SO 
coupling~\cite{hammar,parek}), they will operate with mixed spin states, rather than with the 
originally proposed~\cite{datta90} pure ones.  The full quantum coherence of spin states can be 
preserved only in single-channel wires.   

\acknowledgments
I thank E. I. Rashba and J. Fabian for enlightening discussions.

%********************references***********************************

%*****************************************************************

\end{document}